\documentstyle[prl,aps]{revtex}
\begin{document}
\draft
\twocolumn[\hsize\textwidth\columnwidth\hsize\csname @twocolumnfalse\endcsname

\title{
Selection of the scaling solution in a cluster coalescence model
}
\author{Daniel Kandel\cite{email}}
\address{Department of Physics of Complex Systems,\\
Weizmann Institute of Science, Rehovot 76100, Israel}
\maketitle
\begin{abstract}
The scaling properties of the cluster size distribution of a system of
diffusing clusters is studied in terms of a simple kinetic mean field model.
It is shown that a one parameter family of mathematically valid scaling 
solutions exists. Despite this, the kinetics reaches a unique scaling
solution independent of initial conditions. This selected
scaling solution is marginally physical; i.e., it is the borderline solution
between the unphysical and physical branches of the family of solutions.
\end{abstract}
]
\input epsf

\narrowtext
%
%
Phenomena of diffusion, aggregation and coalescence of clusters of particles 
occur in various scientific areas \cite{general}. 
Of particular interest to us is the relevance of these phenomena
to the physics of thin films. In recent years,
material scientists have been able to observe \cite{mat}
clusters of atoms or vacancies diffusing
on thin film surfaces. The diffusion and coalescence of these clusters
affect the morphology of the film, which in turn affects its electrical
properties and the possibilities to use it in the fabrication of electronic
devices.

In many systems, the time-dependent cluster-size
distribution exhibits a scaling behavior, which 
has been the subject of several existing studies
\cite{meakin,shol}. The
theoretical studies start from a kinetic model of the diffusing clusters,
and then determine the scaling exponents 
by making a scaling ansatz. Finally, the scaling function is found
by solving the kinetic model numerically starting from some
initial cluster size distribution.

The purpose of this work is to investigate the properties of the scaling
function. In particular, the questions answered in this Letter are:
Does the scaling function depend on initial conditions? Starting from 
a scaling ansatz, is there a unique solution for the scaling function?
The answers to both questions turn out to be {\em no}. We show,
on the one hand, that simulations of a cluster coalescence model
starting from different initial cluster size distributions
approach the same scaling function in the long time limit. On the
other hand, solving the equation for the scaling function that results
from a scaling ansatz, we obtain a one parameter family of scaling
functions. This implies
that out of all the solutions for the scaling function, one is 
{\em selected dynamically}. A selection criterion is proposed and verified
numerically.
The applicability of this selection criterion to other models,
which exhibit a scaling behavior,
will be studied in future work.

%
%
The simplest model that describes diffusion and coalescence of clusters
is the model of Meakin \cite{meakin},
where D-dimensional spherical clusters diffuse in 
a d-dimensional space with diffusion constants, ${\cal D}(s)$, that
depend on the cluster size, $s$ (the
number of particles in the cluster). When two
clusters with $s_1$ and $s_2$ particles
touch, they merge irreversibly and form a new spherical cluster
of size $s=s_1+s_2$. Thus, the cluster size distribution depends
on time, and the average cluster size growth is limited only by the
finite number of particles in the system.
Following Meakin, we assume an algebraic decay of
the cluster diffusion constant with size:
\begin{equation}
{\cal D}(s)\sim s^{-\zeta}~~.
\label{dc}
\end{equation}

In the present work we use
$D=d=2$. This is the case of two-dimensional clusters diffusing on 
a two-dimensional surface, appropriate for diffusion of islands (or voids) of
atoms on the surface of a thin film, for example. 
In this case, 
it is possible to show that the cluster diffusion constant indeed 
decays algebraicly with cluster size. The different values
of the exponent $\zeta$ correspond
to various microscopic mechanisms responsible for island diffusion
\cite{ted}. For example, $\zeta=3/2$ when
mass transport on the surface is dominated by diffusion
of atoms along island boundaries. If, however,
mass transport is dominated by exchange of atoms between
islands and nearby terraces, one obtains $\zeta=1$ or
$\zeta=1/2$. When this exchange process is fast compared
with the diffusion of atoms on the terraces, $\zeta=1$.
The case $\zeta=1/2$ is obtained in the opposite limit,
when surface diffusion is fast compared with the exchange kinetics.

To analyze the scaling behavior of such a system, we adopt a mean field
approximation due to Smoluchowski \cite{smol}. In this approach,
the time-dependent density of clusters of size $s$, $\rho(s,t)$,
is assumed to obey the following rate equation:
\begin{eqnarray}
\frac{\partial\rho(s,t)}{\partial t}&=&\sum_{s'=1}^{s-1}
{\cal D}(s')\rho(s',t)\rho(s-s',t) \nonumber\\
&-&\sum_{s'=1}^{\infty}\left[{\cal D}(s)+{\cal D}(s')\right]
\rho(s,t)\rho(s',t)~~.\label{smol}
\end{eqnarray}
The r.h.s. of this equation consists of a gain term due to
the coalescence of clusters of sizes $s'$ and $s-s'$ into a cluster
of size $s$, and a loss term due to the coalescence
of clusters of size $s$ with clusters of all sizes.

%
%
Simulations of various versions of this model have shown 
(see, e.g. \cite{meakin}) that, 
in the long time limit, the density of clusters obeys the scaling
relation 
\begin{equation}
\rho(s,t)=t^{-\alpha} f\left(\frac{s}{\bar s(t)}\right)~~,~~~~~
{\bar s(t)}\sim t^{\beta}~~,
\label{ansatz}
\end{equation}
where ${\bar s(t)}$ is the time dependent average cluster size,
and the exponents $\alpha$ and $\beta$ depend on the form of
the diffusion constant: $\alpha=2/(\zeta+1)$ and $\beta=1/(\zeta+1)$.
These values of the exponents can be easily verified by substituting
the scaling ansatz (\ref{ansatz}) in Eq.\ (\ref{smol}) (see below). 

Here we show that the Smoluchowski equation together with the scaling
ansatz are not sufficient to find the scaling function $f$, since
they lead to a one parameter family of scaling functions. To show this,
we start from the generalized scaling ansatz
\begin{equation}
\rho(s,t)=g(t)f\left(\frac{s}{\bar s(t)}\right)~~.
\label{gansatz}
\end{equation}
We substitute this form of $\rho(s,t)$ into Eq.\ (\ref{smol}), and
change variables from $s$ and $t$ to $u=s/\bar s(t)$ and $t$. 
We also replace summation over $s$ by integration over $u$.
This is justified, since in
the long time limit $\bar s(t)$ is very large. Therefore, when $s$
changes by one, the change in $u$ is much smaller than one. 
These algebraic manipulations together with the functional
form of the diffusion constant, ${\cal D}(s)=K/s^\zeta$, lead to
the following equation:
\begin{eqnarray}
\frac{g'(t)}{Kg(t)^2}\bar s^{\zeta-1}f(u)-\frac{\bar s}{Kg(t)}
\frac{d\bar s}{dt}uf'(u)&&\nonumber\\
=\int_0^u \frac{1}{u'^\zeta}f(u')f(u-u')du' &&\label{eq1}\\
-\int_0^\infty \left(\frac{1}{u^\zeta}+\frac{1}{u'^\zeta}\right)
f(u)f(u')du'&~~,&\nonumber
\end{eqnarray}
where $g'(t)$ and $f'(u)$ are the first derivatives of $g(t)$ and $f(u)$,
respectively. 

Now we take into account the conservation of the total number of 
particles in the clusters, $\Theta$: $\sum_{s=1}^\infty s\rho(s,t)=\Theta$,
where $\Theta$ does not depend on time. Using the scaling ansatz 
(\ref{gansatz}), we obtain a relation between $g(t)$ and $\bar s(t)$:
\begin{equation}
g(t)=\frac{\Theta}{\int_0^\infty uf(u)du}~\frac{1}{\bar s^2}~~.
\label{cons}
\end{equation}
This result can now be combined with Eq.\ (\ref{eq1}) to get the
following equation for the scaling function $f$:
\begin{eqnarray}
-\frac{1}{K\Theta}\bar s^\zeta\frac{d\bar s}{dt}
\int_0^\infty u'f(u')du'\left[2f(u)+uf'(u)\right]
&&\nonumber\\
=\int_0^u \frac{1}{u'^\zeta}f(u')f(u-u')du' &&\label{eq2}\\ 
-\int_0^\infty \left(\frac{1}{u^\zeta}+\frac{1}{u'^\zeta}\right)
f(u)f(u')du'&~~.&\nonumber
\end{eqnarray}

The r.h.s. of Eq.\ (\ref{eq2}) depends on $u$, but not on $t$. Hence,
the l.h.s. of the equation cannot depend on time, and we can rewrite
Eq.\ (\ref{eq2}) in the form
\begin{eqnarray}
\mu
\int_0^\infty u'f(u')du'\left[2f(u)+uf'(u)\right]=
&&\nonumber\\
-\int_0^u \frac{1}{u'^\zeta}f(u')f(u-u')du'+ &&\label{scfeq}\\ 
\int_0^\infty \left(\frac{1}{u^\zeta}+\frac{1}{u'^\zeta}\right)
f(u)f(u')du'&~~,&\nonumber
\end{eqnarray}
where $\mu\equiv \bar s^\zeta\frac{d\bar s}{dt}/K\Theta$ is 
a constant that does not depend on $t$ or $u$. This immediately implies
that $\bar s\sim t^{1/(\zeta+1)}$, and therefore $\beta=1/(\zeta+1)$
as stated above. Since $g\sim\bar s^{-2}$, the value of the exponent
$\alpha$ is also confirmed.
\begin{figure}[h]
  \epsfxsize=95mm
  \centerline{\epsffile{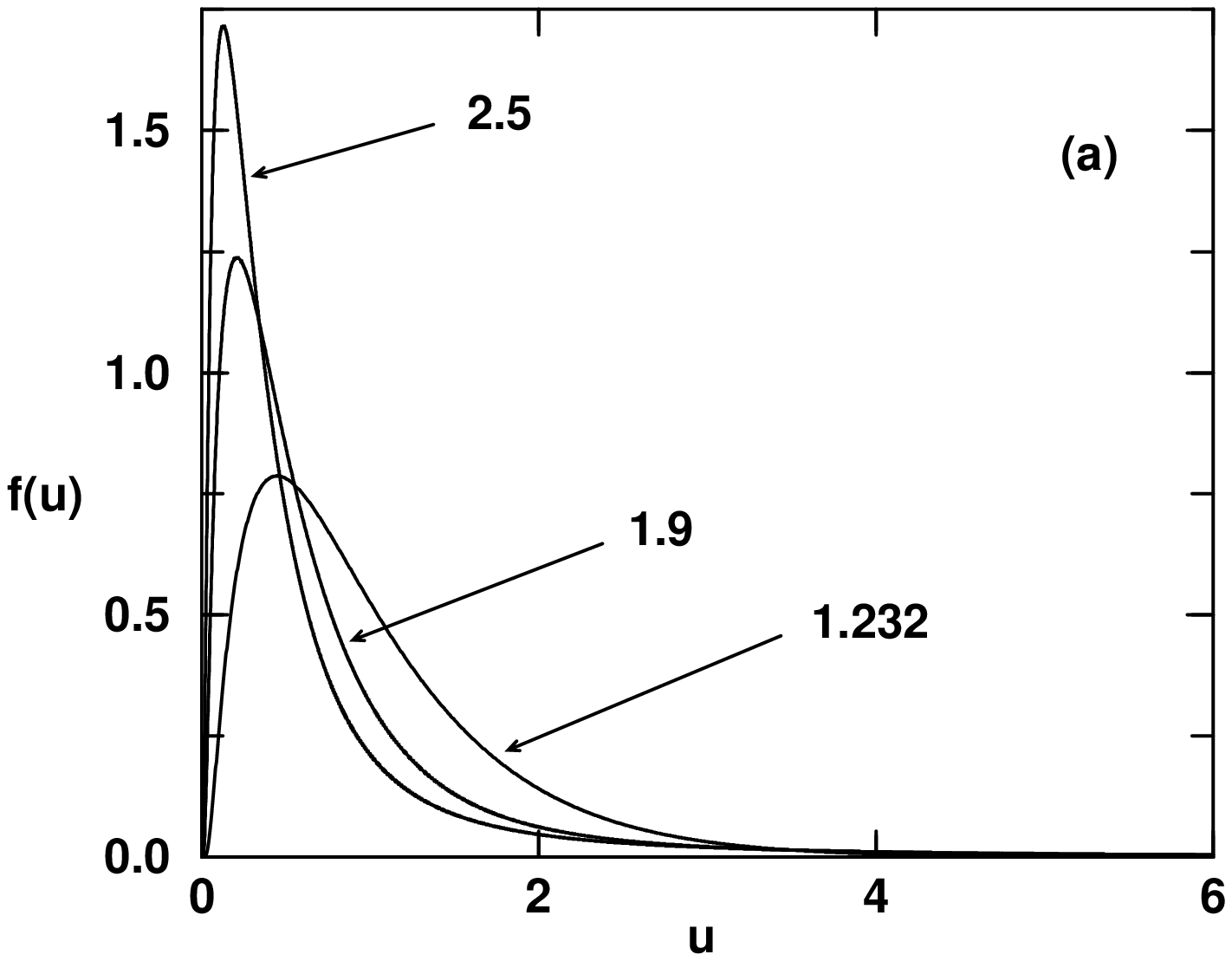}}
  \epsfxsize=95mm
  \centerline{\epsffile{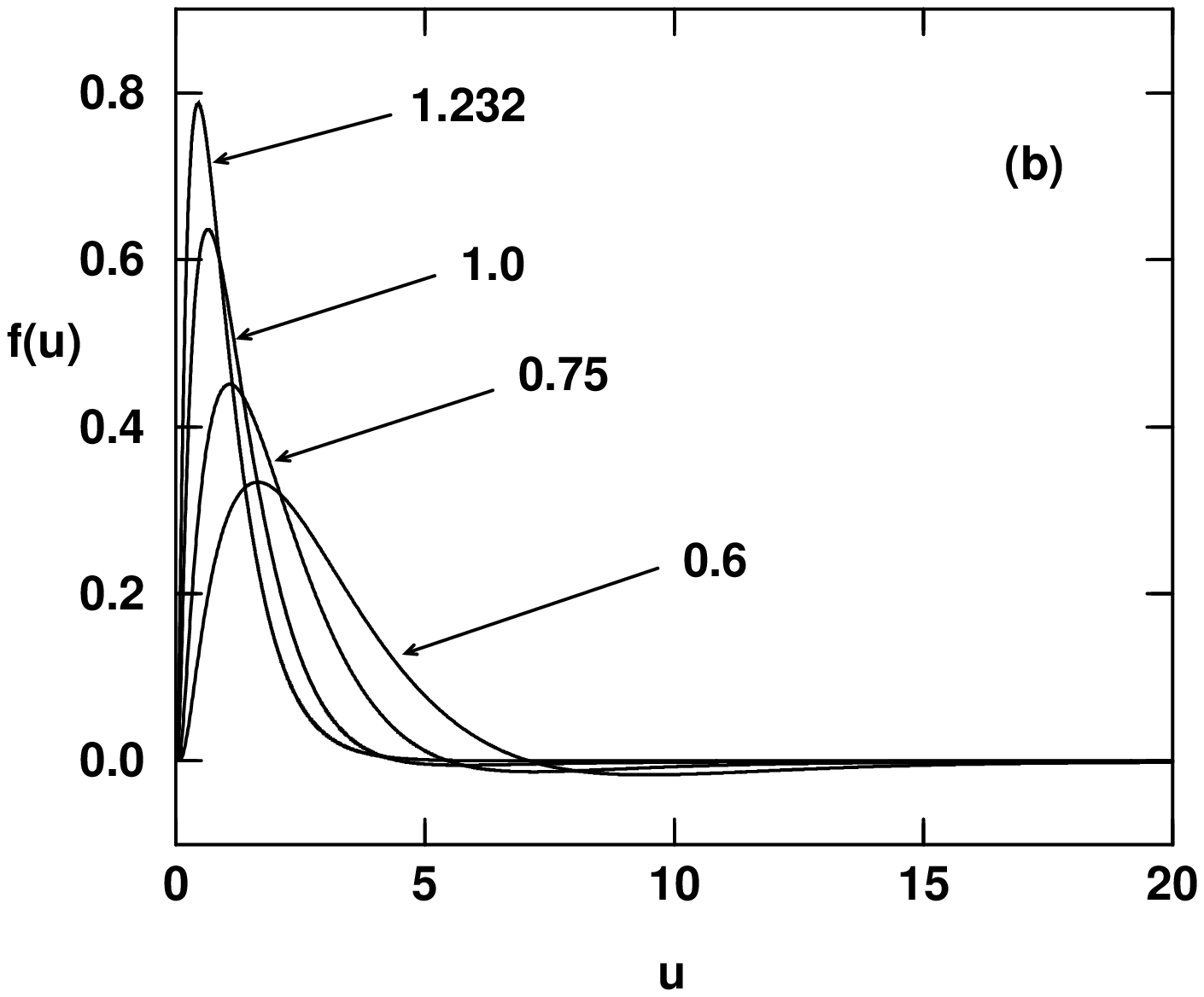}}
\caption{Numerical solutions of the equation for the scaling function with
$\zeta=1/2$. The different curves represent solutions with various
values of $\mu$ (the number near the curves):
(a) $\mu\geq\mu_0$ and (b) $\mu\leq\mu_0$.}
\label{scalf}
\end{figure}

In addition, we can now find the scaling
function $f(u)$ by solving Eq.\ (\ref{scfeq})
with the appropriate boundary condition. In this case, it is an
integral condition that we derive 
from the definition of the average cluster size,
$\bar s\equiv \sum_{s=1}^\infty s\rho(s,t)/\sum_{s=1}^\infty\rho(s,t)$.
Using the scaling ansatz (\ref{gansatz}) in this definition, we obtain
the integral condition
\begin{equation}
\int_0^\infty f(u)du=\int_0^\infty uf(u)du~~.
\label{bc}
\end{equation}

Before we actually solve for the scaling function $f$, it is important 
to note that
one can multiply any solution of Eq.\ (\ref{scfeq}) by a constant
to get a new solution that satisfies the same integral condition. To
eliminate this trivial (and physically meaningless) freedom, we enforce
another integral condition:
\begin{equation}
\int_0^\infty f(u)du=1~~.
\label{bc2}
\end{equation}

%
%
Interestingly, the constant $\mu$ is a free parameter in the problem 
of finding the scaling function as defined above. Our
aim now is to show that there is a range of values of the constant $\mu$,
for which there are legitimate solutions of Eq.\ (\ref{scfeq}) that
obey the integral conditions (\ref{bc}) and (\ref{bc2}). They are all valid
scaling functions for this system.

We solved the equation for the scaling function numerically for $\zeta=1/2$,
using a method that will be outlined elsewhere \cite{else}, and a few
examples of solutions are plotted in Fig.\ \ref{scalf}.
We found that there is
a special value of $\mu$, which we denote by $\mu_0\approx 1.232$, such that
for $\mu\geq\mu_0$ the solutions are perfectly valid. Three examples
of such solutions with $\mu=1.232$, $1.9$ and $2.5$ are shown in 
Fig.\ \ref{scalf}(a).
The solutions
with $\mu<\mu_0$, on the other hand, are unphysical, since $f(u)<0$ 
in these cases in
a range of values of $u$. This can be seen in Fig.\ \ref{scalf}(b)
in the cases $\mu=0.6$, $0.75$ and $1.0$. 
In fact, for even smaller values of $\mu$, the scaling
function develops oscillations, and its limiting behavior at large values
of $u$ becomes ill-defined.
We conclude that there is a one parameter family
of valid and physical scaling functions which correspond to $\mu\geq\mu_0$.

%
%
Our question now is which of these scaling functions are actually reached
by the physical system? This is of course a question about the kinetics
of the system, and in order to answer it we should solve the kinetic
Smoluchowski equation (\ref{smol}) starting from different initial
cluster size distributions. This turns out to be a difficult task that 
consumes enormous amounts of computer time, since the amount of computer
time required for every time step is proportional to the {\em square} of the
number of possible values of cluster sizes.

To circumvent this difficulty, we used a much more efficient {\em simulation
method}. Each simulation started 
with a set of $10^6$--$10^7$ clusters of various sizes
picked according to the chosen initial cluster size distribution. At each
time step a pair of clusters were picked at random. Let us denote their
sizes by $s_1$ and $s_2$. The two clusters were merged into a single
cluster of size $s_1+s_2$ with probability $P(s_1,s_2)$ proportional to
${\cal D}(s_1)+{\cal D}(s_2)$. The clock was then advanced by
$\delta t\sim 1/N(t)^2$, where $N(t)$ is the total number of clusters
in the system at time $t$.
This process was repeated many times thus leading to an evolution of
the cluster size distribution.
It will be shown elsewhere \cite{else}
that the evolution of the {\em average} number
of clusters of size $s$, induced by this kinetic model, follows the
Smoluchowski equation. This procedure is much more efficient than
a simple integration of Eq.\ (\ref{smol}), and the only price 
we have to pay is the statistical error induced by the stochastic nature
of the model.

\begin{figure}[h]
  \epsfxsize=95mm
  \centerline{\epsffile{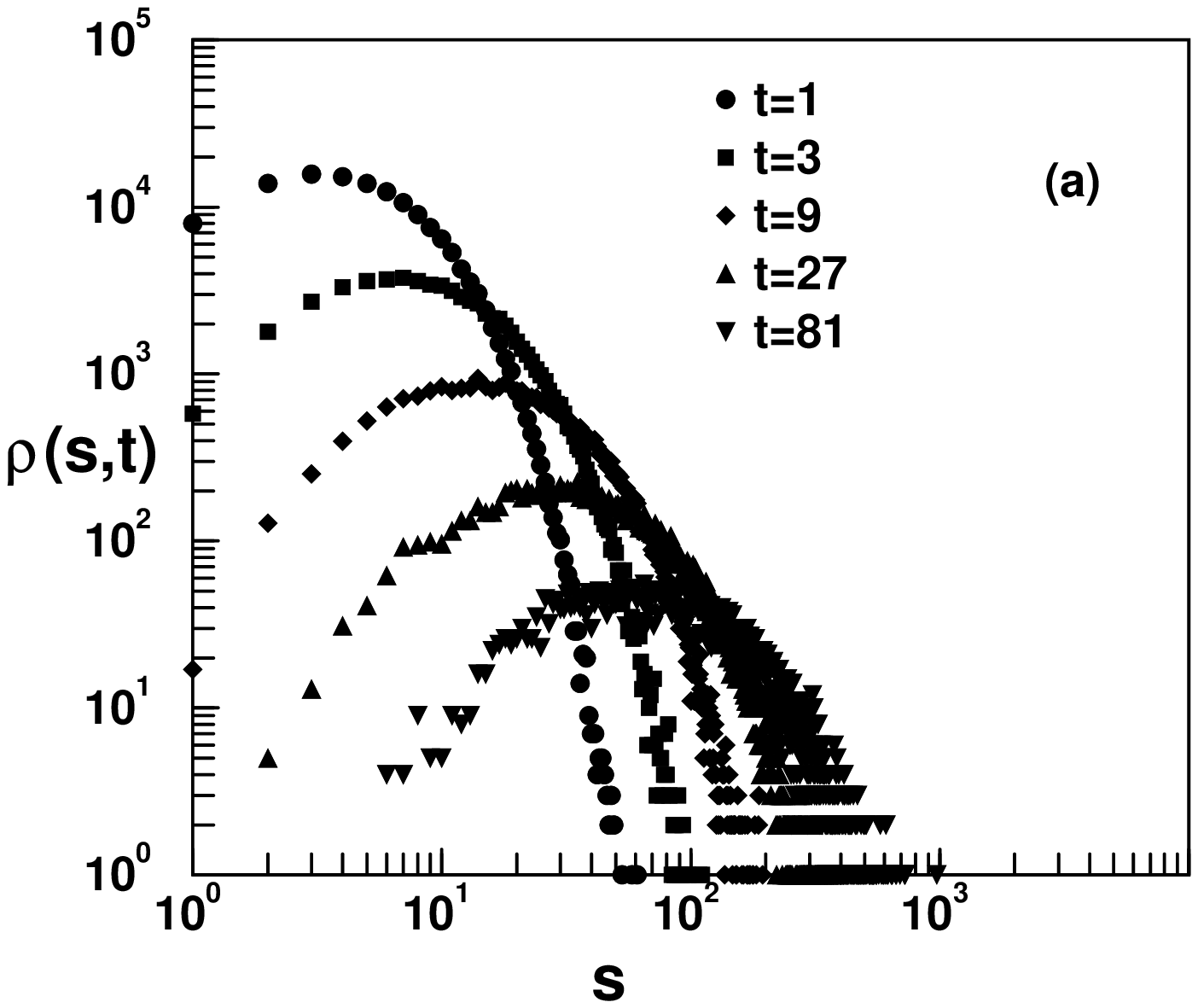}}
  \epsfxsize=95mm
  \centerline{\epsffile{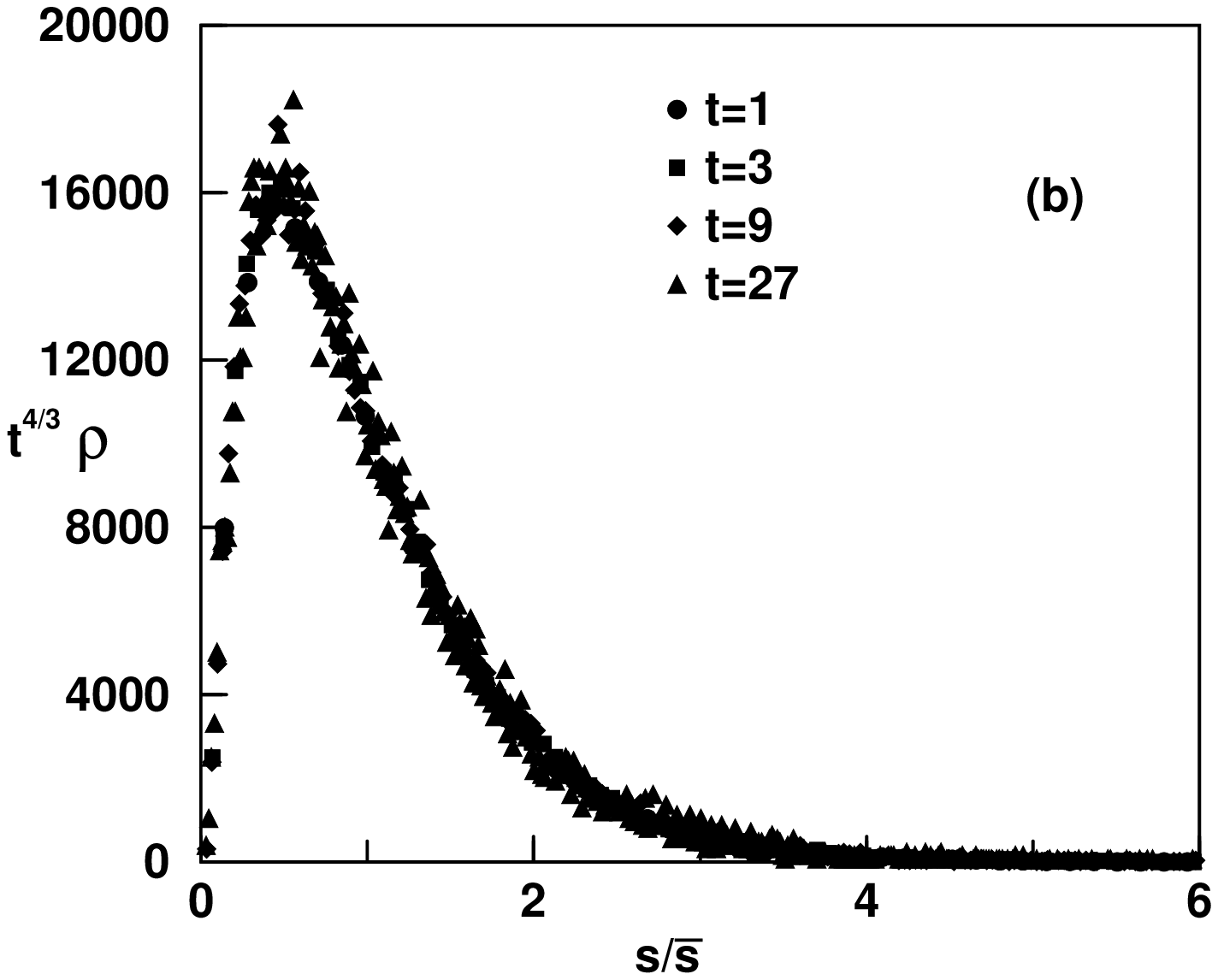}}
\caption{
(a) A log-log plot of
island size distributions resulting from the simulations with
$\zeta=1/2$, at various stages of the evolution of the system.
(b) The scaled island size distributions of (a) form a single curve
to a very good accuracy.
}
\label{rho}
\end{figure}
The simulations were carried out for the case $\zeta=1/2$
(results for other values of $\zeta$ will be presented elsewhere
\cite{else}),
starting from several different initial
cluster size distributions. For example, we started from clusters which were
all of size 25, and also from a uniform distribution of clusters between the
sizes 1 and 50. All the simulations we did exhibited a scaling behavior,
and the scaling function was independent of the initial distribution to
the degree of accuracy of the results. In Fig.\ \ref{rho}(a) we show the
cluster size distributions, at different times, 
obtained from one of the simulations starting
from a set of $10^6$ clusters, all of size 1. We see from the figure
that the most probable cluster size as well as the width of the distribution
grow with time. Fig.\ \ref{rho}(b) shows $t^\alpha\rho(s,t)$ as a function
of the scaled cluster size $u=s/\bar s(t)$, with $\alpha=4/3$ as deduced
from the scaling analysis for $\zeta=1/2$ (see above). Clearly, there
is excellent data collapse and all the distributions fall on top of
a single curve, which is the scaling function $f(u)$ up to a multiplicative
constant. Examination of the average cluster size as a function of time
shows that
$\bar s(t)\sim t^\beta$ with $\beta=2/3$ as expected from the scaling
analysis for $\zeta=1/2$ (see above).

%
%
\begin{figure}[h]
  \epsfxsize=95mm
  \centerline{\epsffile{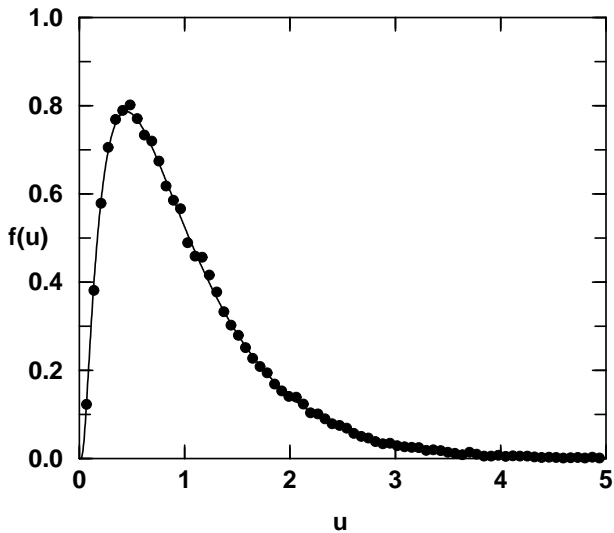}}
\caption{
The scaling function obtained from the simulations is shown as
full circles. The solid line is the solution of the scaling function
equation with $\mu=\mu_0\approx 1.232$.
}
\label{select}
\end{figure}
Our simulation results indicate that the scaling function is not sensitive
to the initial cluster size distribution. This intriguing result implies
that one scaling function (and a unique value of the parameter $\mu$)
is {\em selected dynamically} out of the one parameter family of
possible physical scaling functions. Which one is it? What is the
selected value of $\mu$? The solution to this puzzle can be inferred from
Fig.\ \ref{select}, which shows the scaling function obtained from
simulations (normalized so that $\int_0^\infty f(u)du=1$) as full circles,
and the scaling function for $\mu=\mu_0\approx 1.232$ as a solid line. 
The resemblance of the two functions suggests that the kinetically selected
scaling solution is the one at the border between the physical
and unphysical branches of the family of solutions, i.e., the scaling
function that corresponds to $\mu=\mu_0$. 

One can argue that this selection principle is reasonable, since the
initial cluster size distribution is cut off at some finite size.
Assuming that the tail of the distribution develops gradually, the
system will approach the physically legitimate distribution with the
shortest tail. A close examination of the family of solutions leads
to the conclusion that the scaling function with $\mu=\mu_0$ 
has the shortest tail, since it decays
to zero faster than all the scaling functions with $\mu>\mu_0$.

A similar selection principle was proposed by Stavans et al.
and Segel et al. \cite{stavans}
in relation with the selection of the steady state in the coarsening
of cells in two-dimensional soap froths, and by Maggs et al. \cite{maggs}
in connection with steady state length distributions of living polymers.
This suggests that the selection
principle proposed here may apply to a class of problems. In future work, 
we intend to investigate its
applicability in various non-equilibrium
systems that exhibit a scaling behavior.

This research was supported by grant No. 95-00268 from the 
United States-Israel Binational Science Foundation (BSF), Jerusalem, Israel.
I would like to thank E. Domany, T. L. Einstein, Y. Manassen, D. Mukamel, 
Z. Olami, J. Stavans and J. D. Weeks for discussions.
D. Kandel is the incumbent of the Ruth Epstein Recu
Career Development Chair.


%
%
\end{document}